\documentclass[11pt]{article}

\usepackage{amsmath}
\usepackage{amssymb}
\usepackage{epsfig}

\newtheorem{theorem}{Theorem}[section]

\newcommand{\bea}{\begin{eqnarray*}}
\newcommand{\eea}{\end{eqnarray*}}

\begin{document}


\begin{titlepage}

\renewcommand{\thefootnote}{\alph{footnote}}
\vspace*{-3.cm}
\begin{flushright}

\end{flushright}

\vspace*{0.5cm}

\renewcommand{\thefootnote}{\fnsymbol{footnote}}
\setcounter{footnote}{-1}

{\begin{center} {\Large\bf Partition Identities From Partial
Supersymmetry}

\end{center}}
\renewcommand{\thefootnote}{\alph{footnote}}

\vspace*{.8cm}
{\begin{center} {\large{\sc
                Noureddine~Chair$^{a,b}$
                }}
\end{center}}
\vspace*{0cm}
{\it
\begin{center}
 $^a$Physics Department,
Al al-bayt University, Mafraque, Jordan

Email: n.chair@rocketmail.com\\
\hspace{19mm}nchair@alalbayt.aabu.edu.jo
\end{center} }

\vspace*{0cm}
{\it
\begin{center}
 $^b$The Abdus Salam International Centre For Theoretical Physics,
 1-34014 Trieste,Italy,

Email: chairn@ictp.trieste.it
\end{center} }

\vspace*{1.5cm}

{\Large \bf
\begin{center} Abstract \end{center} }
In the quantum theory, using the notion of partial supersymmetry,
in which some, but not all, operators have superpartners  we
derive the Euler theorem in partition theory. The paraferminic
partition function gives another identity in partition theory with
restrictions. Also an explicit formula for the graded
parafermionic partition function is obtained. It turns out that
the ratio of the former partition function to the latter is given
in terms of the Jacobi Theta function, $\theta_{4}$. The inverted
graded parafermionic partition function is shown to be a
generating function of partitions of numbers with restriction that
generalizes the Euler generating function and as a result we
obtain new sequences of partitions of numbers with given
restrictions.

\vspace*{.5cm}

\end{titlepage}

\newpage

\renewcommand{\thefootnote}{\arabic{footnote}}
\setcounter{footnote}{0}


%

\section{INTRODUCTION \label{sec:SEC-intro}}
In arithmetic quantum theories the spectrum is chosen to be
logarithmic in order to connect quantum mechanics both to number
theory (multiplicative number theory), since the partition
function is related to the Riemann zeta or other Dirichlet series
\cite{donald} and to string theory \cite{i.bakas}. In these
theories  number theoretic identities have been derived and
interpreted. Spector in \cite{spector} introduced a notion of
partial supersymmetry in which some, not all, bosonic states or
operators have superpartners, with this notion again he derived
and interpreted the fermionic and parafermionic thermal partition
functions. In this paper we use the notion of supersymmetry  in a
quantum theory free of a logarithmic spectrum and  connect it to
the additive number theory, \cite{apostol} since our partition
function is related to the Euler generating function for
partitions. In so doing we have derived the Euler theorem from a
fermionic partition function, the theorem says that the number of
partitions of a number $k$ containing odd numbers only equals the
number of partitions of a number $k$ without duplication.
Similarly, from parafermionic partition function we have derived
another identity in partition theory \cite{andrews} which equates
the number of partitions of $k$ in which no part appears more than
$s-1$ times, with the number of partitions of $k$ such that no
part is divisible by $s$. Using this identity we prove Andrews'
result \cite{andrews} in connection with generating functions that
exclude squares and their generalizations \cite{sellers}. In the
graded parafermionic case we have obtained an explicit formula for
the partition function and have shown that it is the inverted
parafermionic partition function that corresponds to the
generalized generating function for partitions in the sense of
Euler since the Euler's generating function of partitions it has
the interpretation of being the inverted graded fermionic
partition function. With this generating function new sequences of
partitions have appeared \cite{sloane}

\section{ PARTITION FUNCTIONS AND THE EULER IDENTITY}
Here we first review the concept of partial supersymmetry
introduced by Spector \cite{spector}, then using this concept we
obtain the fermionic partition function written in terms of the
bosonic partition function and the graded fermionic partition
function and as a result the Euler identity is obtained. The Euler
identity says that the generating functions for the number of
partitions of a given number $k$ into distinct parts and the
number of partitions of $k$ into odd parts are equal. In this
paper we will be considering non-interacting quantum field
theories which are not of a logarithmic spectrum, and so we can
write the bosonic(fermionic) Hamiltonian without the logarithm of
a prime in the form,
\begin{equation}
\label{toto1}
 H_B=\omega\sum_{k=1}^\infty b_k^{\dag}b_k.
\end{equation}
\begin{equation}
\label{toto2}
 H_F=\omega\sum_{k=1}^\infty f_k^{\dag}f_k
\end{equation}
where $b_k^{\dag}(b_k)$ are the bosonic creation(annihilation)
operators respectively and $f_k^{\dag}(f_k)$ are the fermionic
creation (annihilation) operators respectively.The basic idea of
partial supersymmetry is that not all bosons have superpartners,
therefore if we start with a bosonic partition function in which
the Hamiltonian is decomposed into bosonic and fermionic parts and
then using the Witten index \cite{witten} the bosons are cancelled
and we are left with the fermionic partition function.Suppose we
define operators $q_k$ and $c_k$ such that $q_k=(b_k)^2$,
$q_k^{\dag}=(b_k^{\dag})^2$ and $(c_k)^2=(c_k^{\dag})^2=0$, note
that $c_k$ and $c_k^{\dag}$ have the same effect as $b_k$ and
$b_k^{\dag}$ respectively the difference being that they are
square free. Equivalently the bosonic Hamiltonian is
\begin{equation}
\label{toto3}
 H_B=\omega\sum_{k=1}^\infty c_k^{\dag}c_k+\omega\sum_{k=1}^\infty 2q_k^{\dag}q_k.
\end{equation}
Now, since the components of our decomposition of the Hamiltonian,
$H_B$  commute then the trace in the bosonic partition function
decomposes as
\begin{equation}
\label{toto4} {\rm Tr}\left[\exp{-\beta H_B}\right]= {\rm
Tr}\left[\exp{-\beta\omega\sum_{k=1}^\infty c_k^{\dag}c_k}\right]
{\rm Tr}\left[\exp{-\beta\omega\sum_{k=1}^\infty
2q_k^{\dag}q_k}\right],
\end{equation}
where the first term on the right of this equation is the
fermionic partition function and the second term is the bosonic
partition function. Now adding the term $\omega\sum_{k=1}^\infty
2f_k^{\dag}f_k$ to the original Hamiltonian and then using the
witten index
\begin{eqnarray}{\rm
Tr}\left[(-)^F\exp{-\beta(H_B+H_F)}\right]=1,\nonumber
\end{eqnarray}
where$F$ is the fermion number operator with eigenvalues 0 or 1.
Note that partial supersymmetry here means we have fermionic
superpartners for some of the bosonic creation operators ( $q_k$
and $q_k^{\dag}$) but not others ($c_k$ and $c_k^{\dag}$) so one
does not grade the $c_k$ partition function . Therefore we obtain
the following factorization identity;
\begin{eqnarray}
\label{toto5}
 Z_f(\beta)&=&{\rm
Tr}\left[\exp(-\beta\omega\sum_{k=1}^\infty
c_k^{\dag}c_{k})\right]= {\rm
Tr}\left[(-1)^F\exp{-\beta\omega\sum_{k=1}^\infty
(c_k^{\dag}c_k+2q_k^{\dag}q_k+2f_k^{\dag}f_k)}\right]\nonumber\\
&=& {\rm Tr}\left[(-)^F\exp{-\beta(H_B+2H_F)}\right].
\end{eqnarray}
The bosonic and the fermionic partition functions in equation
(\ref{toto5}) are computed in the Fock space of states for both
bosons and fermions using the expressions for $H_B$ and $H_F$
given by equation (\ref{toto1}) and equation (\ref{toto2})
respectively, and so we have,
\begin{eqnarray}
\label{toto6}
Z_b(\beta)&=& {\rm Tr}\left[\exp{-\beta H_B}\right]=
{\rm Tr}\left[\exp{-\beta\omega\sum_{k=1}^\infty
b_k^{\dag}b_k}\right]=\prod_{k=1}^\infty\sum_{i=0}^\infty x^
{ik}=\prod_{k=1}^{\infty}\frac{1}{1-x^k},
\end{eqnarray}
where we have set $x=\exp{-(\beta\omega)}$, one can easily check
that the partition function for equation (\ref{toto3}) and that
for $H_{B}$ equation (\ref{toto1}) agree. The computation in the
fermionic case is similar except that due to the Fermi-Dirac
statistics $i$ takes the values $0$ and $1$ therefore
\begin{eqnarray}
\label{toto7}Z_f(\beta)&=&{\rm
Tr}\left[\exp{-\beta\omega\sum_{k=1}^\infty
c_k^{\dag}c_k}\right]=\prod_{k=1}^\infty\sum_{i=0}^1 x^
{ik}=\prod_{k=1}^{\infty}({1+x^k}).
\end{eqnarray}
The remaining partition function to be computed is the graded
fermionic partition \begin{eqnarray} {\rm
Tr}\left[(-)^F\exp{-\beta(2H_F)}\right] \nonumber.
\end{eqnarray} Since the fermion number (the eigenvalue of $F$) is either $0$
or $1$, then if we denote this partition by $\Delta_F(\beta)$ we
have
\begin{eqnarray}
\label{toto8}\Delta_F(2\beta)&=&{\rm
Tr}\left[(-1)^F\exp(-2\beta\omega\sum_{k=1}^{\infty}
c_{k}^{\dag}c_{k})\right]=\prod_{k=1}^{\infty}\sum_{i=0}^{1}(-1)^{i}
x^{ik}=\prod_{k=1}^{\infty}(1-x^{2k}).
\end{eqnarray}
Finally, the factorization identity gives

\begin{eqnarray}
\label{toto9}\prod_{k=1}^{\infty}({1+x^k})&=&
\prod_{k=1}^{\infty}\frac{(1-x^{2k})}{({1-x^k})}=
\prod_{k=1}^{\infty}\frac{(1-x^{2k})}{(1-x^{2k})(1-x^{2k-1})}=
\prod_{k=1}^{\infty}\frac{1}{(1-x^{2k-1})}.
\end{eqnarray}
This is a well known theorem due to Euler \cite{apostol} which
says that the generating functions for the number of partitions of
a given number $k$ into distinct parts and the number of
partitions of $k$ into odd parts are equal. Here we have obtained
this equality using partial supersymmetry. Next we will write down
the partitions function for the bosonic and fermionic simple
harmonic oscillators respectively, then connect them through
partial supersymmetry and as a result we get a well known identity
for $\cosh{x}$ in terms of an infinite  product. The Hamiltonian
for the bosonic(fermionic) simple harmonic oscillator are
$H_B=\omega (b{\dag}b+{1/2}) $( $H_F=\omega (f^{\dag}f-{1/2}))$
respectively and hence their partition functions are,
\begin{eqnarray}
\label{toto10} Z_b(\beta) &=&{\rm
Tr}\left[\exp(-{\beta\omega}(b^{\dag}b+{1/2}))\right] \nonumber\\
&=& \sum_{n=0}^{\infty}\exp(-\beta\omega(n+{1/2}))=
\frac{1}{2\sinh(\beta\omega/2)}.
\\\label{toto11}
Z_f(\beta) &=&{\rm
Tr}\left[\exp(-\beta\omega(f^{\dag}f -{1/2}))\right]\nonumber\\
&=&\sum_{n=0}^{1}\exp\left[- \beta\omega(n-  {1/2})\right] =
2(\cosh(\beta\omega/2)),
\end{eqnarray}
The other piece we need to compute is the graded fermionic
partition function $$\Delta_{F}(2\beta)= {\rm
Tr}\left[(-1)^{F}\exp(-\beta\omega(f^{\dag}f-{1/2}))\right)]$$
which is simple to calculate since the eigenvalues of the fermion
number operator $F$ are $0$ and $1$ and so we have
$\Delta_{F}(2\beta)=2\sinh(\beta\omega/2)$ since the infinite
product representation of  sine hyperbolic is
$\sinh(x)=x\prod_{k=0}^{\infty}(1+\frac{x^{2}}{k^{2}\pi^{2}})$.
Then the factorization identity (\ref{toto5}) gives
\begin{eqnarray}
\label{toto12}
Z_f(\beta)&=&2\cosh(\beta\omega)/2)=\frac{2\sinh(\beta\omega)}{2\sinh(\beta\omega/2)}=
2\prod_{k=1}^{\infty}\frac{(1+\frac{(\beta\omega)^{2}}{k^{2}\pi^{2}})}{(1+\frac{(\beta\omega)^{2}}{4(k)^{2}\pi^{2}})}=
2\prod_{k=1}^{\infty}{(1+\frac{(\beta\omega)^{2}}{(2k+1)^{2}\pi^{2}})},
\end{eqnarray}
which is exactly the infinite product representation of
$(\cosh(\beta\omega)/2)$ therefore we see that the results one
obtain from partial supersymmetry depends very much on the
Hamiltonian used. If one is dealing with a theory with a
logarithmic spectrum then partial supersymmetry gives a proof of a
number theoretic identity that connects the zeta function to the
modulus of Mobius inversion function \cite{spector}. Here partial
supersymmetry it gives the generating function proof of the Euler
theorem in which the number of partitions of $k$ into distinct
parts equals the number of partitions of $k$ into odd parts. Also
in the case the Hamiltonian is that of a harmonic oscillator we
have derived a well known  identity in hyperbolic trigonometry
$\cosh(\beta\omega)/2)=\frac{\sinh(\beta\omega)}{\sinh(\beta\omega/2)}$.
From which we obtain the infinite product representation of
$\cosh{x}$ knowing that of $\sinh(x)$.
\section{PARAFERMIONIC PARTITION FUNCTIONS AND PARTITIONS WITH RESTRICTIONS}
The natural generalization to the previous section, in which the
fermionic partition function was factorized as a product of graded
fermionic partition function $\Delta(2\beta)$ times the bosonic
partition function $Z_b(\beta)$, would be to consider parafermions
of order s. As fermions are of order $2$ therefore, just like the
first factorization identity (\ref{toto5}) we will have the
following second factorization identity
\begin{equation}
\label{toto13} Z_s(\beta)= {\rm
Tr}\left[(-)^F\exp{-\beta(H_B+sH_F)}\right] ,
\end{equation}
where the Hamiltonian $H_B$ is constructed out of certain
operators $\chi_k$ and $r_k$ such that
$(\chi_k)^s=(\chi_k{\dag})^s=0$ but no lower powers vanish as
operators,i.e., these are the parafermionic operators and the
bosonic operators are $ (r_k)^s= (b_k)^s$, $(r_k\dag)^s=
(b_k\dag)^s$ thus
\begin{equation}
\label{toto14}
 H_B=\omega\sum_{k=1}^\infty \chi_k^{\dag}\chi_k+\omega\sum_{k=1}^\infty sr_k^{\dag}r_k.
\end{equation}
 By the Witten index, the parafermionic partition reads,
\begin{equation}
\label{toto15} Z_s(\beta)= {\rm Tr}\exp(-\beta(H_{s}))={\rm
Tr}\left[(-)^{F}\exp(\beta(H_B+sH_F))\right]
\end{equation}
 where
$H_{s}=\omega\sum_{k=1}^{\infty} \chi_{k}^{\dag}\chi_k$. The
parafermionic partition function is a sort of truncated bosonic
partition since the term $x^{sk}$ and higher terms are not present
and so $Z_{s}(\beta)=\prod_{k=1}^{\infty}(1+ x^{k} + x^{2k}+
\cdots + x^{(s-1)k})$ and therefore, using the second
factorization identity, we obtain

\begin{eqnarray}
\label{toto16} Z_{s}(\beta)&=&\prod_{k=1}^{\infty}(1+ x^{k} +
x^{2k}+ \cdots +
x^{(s-1)k})\nonumber\\
&=&\prod_{k=1}^{\infty}\frac{(1-x^{sk})}{({1-x^k})}.
\end{eqnarray}
This is exactly an identity in the theory of partitions
\cite{andrews} which says that the the generating function for the
number of partitions of $k$ in which no parts occur more than $
s-1 $ times equals  the generating function for the number of
partitions of $k$ such that no parts is divisible by $s$. This is
understood as we have eliminated terms of the form
$\prod_{k=1}^{\infty}(1-x^{sk})$ from the bosonic partition
function which in turn is the generating function for the number
of partitions of $k$ without restrictions. Therefore this well
known result in partition theory with restriction is obtained from
partial supersymmetry and our Hamiltonians $H_s$, $H_B$ and $H_F$.
Before proving some theorems in the theory of partitions using the
second factorization identity, let us first consider the following
simple example, take a parafermion of order three so its partition
function by the second factorization identity is
$Z_{3}(\beta)=\prod_{k=1}^{\infty}(1+ x^{k} +
x^{2k})=\prod_{k=1}^{\infty}\frac{(1-x^{3k})}{({1-x^k})}$, the
right hand side of this identity can be simplified to give
\begin{equation}
\label{toto17}
Z_{3}(\beta)=\prod_{k=1}^{\infty}\frac{1}{(1-x^{3k-2})}\frac{1}{(1-x^{3k-1})},
\end{equation}
where we have used the identity
$\prod_{k=1}^{\infty}(1-x^{3k-2})(1-x^{3k-1})(1-x^{3k})=\prod_{k=1}^{\infty}({1-x^k})$
the right hand of equation (\ref{toto17}) is the number of
partitions of $k$ into parts prime to 3 and so equals to the
number of partitions of $k$ in which each part occurs at most two
times. Using Maple for the product in equation (\ref{toto17}) one
obtain the following
series\begin{center}$1+x+2x^{2}+\cdots+9x^{7}+13x^{8}+\cdots+1225x^{30}+\cdots$
.\end{center} In terms of partitions this means for example that
the number of partition of the number 7 which are prime to 3 is 9
because
\begin{center}$7=5+2=5+1+1=4+2+1=4+1+1+1=2+2+2+1=2+2+1+1+1=2+1+1+1+1+1=1+1+1+1+1+1+1$.\end{center}
The sequence of partitions numbers corresponds to the coefficients
in the series therefore this sequence is,
\begin{center}$1,1,2,\cdots,9,13,\cdots,1225,\cdots$,\end{center}
this sequence coincides with the sequence with reference number
A00726 in the On-Line encyclopedia of integer sequences
\cite{sloane}.

Next using the second factorization equation (\ref{toto15}) we
will give a different proof for the following three theorems. one
theorem is on Andrews' result \cite{andrews} in connection with
generating functions that exclude squares and the other two
theorems  are their generalizations \cite{sellers}. The Andrews'
result stated in \cite{sellers} by the following theorem;

\begin{theorem}{Let $P_1(n)$ be the number of partitions of $n$ in
which each $k$ appears at most $k-1$ times and let $P_2(n)$ be the
number of partitions of $n$ with no squares part. Then,
$P_1(n)=P_2(n)$}.
\end{theorem}
Proof. The generating function for $P_1(n)$ is nothing but the
parafermionic partition function of order $k$,i.e.,
\begin{equation}
\label{toto18} \sum_{n=0}^\infty
P_1(n)x^{n}=\prod_{k=1}^{\infty}(1+ x^{k} + x^{2k}+ \cdots +
x^{(k-1)k})=\prod_{k=1}^{\infty}\frac{(1-x^{k^{2}})}{({1-x^k})},
\end{equation}
the right hand side is the generating function for the number of
partitions in which no square parts are present and so equals to
$\sum_{n=0}^\infty P_2(n)x^{n}$ and hence $P_1(n)=P_2(n)$.

The generalization considered in \cite{sellers} is to exclude
polygonal numbers or $r$-gons as parts  where the general
polygonal number or the $k^{th} n$-gonal number is given by
$p_k^n=\frac{1}{2}k \left[(n-2)k-(n-4)\right]$ setting $n=4$ this
equation gives square numbers $k^{2}$ and $n=5$ gives the
$k^{th}-pentagonal$ numbers $\frac{1}{2}k (3k-1)$, etc.
\begin{theorem}{Let $r\geq2$ be a fixed integer . Let $P_3(n,r)$ be the number of partitions of $n$
in which each $k$ appears at most $(r-1)(k-1)$ times and let
$P_4(n,r)$ be the number of partitions of $n$ where no $2r$-gons
can be used as parts.Then, $P_3(n,r)=P_4(n,r)$}.
\end{theorem}
Proof. The generating function for $P_3(n,r)$ is the parafermionic
partition function of order $(r-1)(k-1)+1$ and so we have;
\begin{equation}
\label{toto19} \sum_{n=0}^\infty
P_3(n,r)x^{n}=\prod_{k=1}^{\infty}(1+ x^{k} + x^{2k}+ \cdots +
x^{k(r-1)(k-1)})=\prod_{k=1}^{\infty}\frac{(1-x^{k[(r-1)(k-1)+1]})}{({1-x^k})},
\end{equation}
to complete the proof we simply note that the term
$k[(r-1)(k-1)+1]$ can be rewritten as $k
\left[(r-1)k-(r-2)\right]=\frac{1}{2}k\left[(2r-2)k-(2r-4)\right]$
which is the $2r$-gons so the right hand side of the above
equation generates partitions whose parts are free of $2r-gons$
and hence $P_3(n,r)=P_4(n,r)$. Note that if we set $r=2$ we obtain
the previous result in which no square parts are present in the
partitions.
The next results obtained in \cite{sellers} is to
exclude $2r+1$-gons parts from partitions and is stated as
follows;
\begin{theorem}{Let $P_5(n,r)$ be the number of partitions of $n$ in
which each the part $2k-1$ $(k\geq1)$ appears at most
$(2r-1)(k-1)$ times (and the frequency of even parts is
unbounded). let $P_6(n,r)$ be the number of partitions of $n$ in
which no odd-subscribed $2r+1$-gons can be used as parts. Then,
for all non-negative $n$ $P_5(n,r)=P_6(n,r)$ }.
\end{theorem}
Proof. The generating function $P_5(n,r)$ from its definition is a
product of even bosonic partition
$\prod_{k=1}^{\infty}\frac{1}{({1-x^{2k}})}$ and the odd
parafermionic partition function of order $[(2r-1)(k-1)+1]$.
Therefore, similarly to the above proof we have
\begin{eqnarray}
\label{toto20} \sum_{n=0}^\infty
P_{5}(n,r)x^{n}&=&\prod_{k=1}^{\infty}\left(\frac{1}{1-x^{2k}}\right)\left(1+
x^{2k-1} + x^{2(2k-1)}+ \cdots +
x^{(2k-1)(2r-1)(k-1)}\right)\nonumber\\
&=&\prod_{k=1}^{\infty}\left(\frac{1}{1-
x^{2k}}\right)\left(\frac{1- x^{(2k-1)[(2r-1)(k-1)+1]}}{1-
x^{(2k-1)}}\right)\nonumber\\
&=&\prod_{k=1}^{\infty}\left(\frac{1-
x^{(2k-1)[(2r-1)(k-1)+1]}}{1- x^k}\right),
\end{eqnarray}
from the following algebraic identity
\begin{eqnarray}
\label{toto21}
(2k-1)[(2r-1)(k-1)+1]&=&(2k-1)[(2r-1)k-(2r-2)]\nonumber\\
 &=&
\frac{(2r+1-2)}{2}(2k-1)^2-\frac{(2r+1-4)}{2}(2k-1),
\end{eqnarray}
so the term $(2k-1)[(2r-1)(k-1)+1]$ is nothing but the $2r+1$-gons
by definition, therefore $P_5(n,r)=P_6(n,r)$.

\section{GRADED PARAFERMIONIC PARTITION FUNCTIONS AND OTHER IDENTITIES}
We have already seen in section two that the graded  fermionic
partition function $\Delta_F(\beta)={\rm
Tr}\left[(-)^F\exp{-\beta(H_F)}\right]=\prod_{k=1}^{\infty}(1-
x^{k})$  so grading is equivalent to changing the  sign of $x^k$
in the fermionic partition function. Similarly  the graded bosonic
partition function can be obtained by changing the sign of $x^k$
in the bosonic partition function, so if we denote by
$\Delta_B(\beta)$ the graded bosonic partition function then we
have
\begin{eqnarray}
\label{toto22}
\Delta_B(\beta)&=&\prod_{k=1}^{\infty}\frac{1}{(1+x^k)}\nonumber\\
&=& \prod_{k=1}^{\infty}\sum_{i=0}^{\infty}(-1)^{i} x^{ik},
\end{eqnarray}
therefore the right hand of the above equation  splits into even
terms with plus sign in front and the odd terms with a minus sign
in front and hence in terms of operators the graded bosonic
partition function can be written as $\Delta_B(\beta)={\rm
Tr}\left[(-1)^{N_B}\exp{-\beta(H_B)}\right]$, where the operator
$(-1)^{N_B}=\pm1$. Note that the latter operator comes naturally
here in trying to obtain the graded bosonic partition function. It
was introduced in \cite{spector} in connection with the notion of
the bosonic index; it is $+1$ for Fock space states with an even
number of bosonic creation operators and $-1$ for Fock space
states with an odd number of bosonic creation operators. Recall
from the last section that the parafermion partition function is a
truncation of the bosonic partition function and so the graded
parafermion partition function would be the truncation of the
graded bosonic partition function. There are two truncations to
consider depending on the order $s$ of the parafermion,  when it
is even $(-1)^s=+1$ or when it is odd $(-1)^s=-1$. we shall denote
these partition functions by $ \Delta_s^{\pm}(\beta)$. Therefore
using our physical intuition, the graded parafermionic partition
function is obtained from parafermionic partition function simply
by changing $x^k$ to $-x^k$ . Therefore the identity in equation
(\ref{toto16}) with $x^k$ changed to $-x^k$ gives another
mathematical identity,
\begin{equation}
 \label{toto23}
\prod_{k=1}^{\infty}(1- x^{k} + x^{2k}- \cdots +(-1)^{s-1}
x^{(s-1)k})=\prod_{k=1}^{\infty}\frac{(1+(-1)^{s-1}x^{sk})}{({1+x^k})}.
\end{equation}
Rewriting the above identity in operator form and considering
separately  $s$ even and odd, we end up with the following two
formulae,

\begin{eqnarray}
\label{toto24} \Delta_s^{+}(\beta)&=&{\rm
 Tr}\left[(-1)^s\exp{-\beta(H_s)}\right]\nonumber\\
&=&{\rm Tr}\left[(-1)^{N_B}\exp{-\beta(H_B)}\right] {\rm
 Tr}\left[(-1)^F\exp{-\beta(H_F)}\right],\;\mbox\;s\;\mbox{even}
\end{eqnarray}

\begin{eqnarray}
\label{toto25} \Delta_s^{-}(\beta)&=&{\rm
Tr}\left[(-1)^s\exp{-\beta(H_s)}\right]\nonumber\\
&=&{\rm Tr}\left[(-1)^{N_B}\exp{-\beta(H_B)}\right]{\rm
Tr}\left[\exp{-\beta(H_F)}\right],\;\mbox\;s\;\mbox{odd}
\end{eqnarray}
which are the identities obtained in \cite{spector}. Using partial
supersymmetry and the Witten index in the even case, and partial
supersymmetry and the notion of the bosonic index in the odd case.
As a consequence of the identities given in equations
(\ref{toto16}) and (\ref{toto23}), the ratio of the graded
parafermionic partition function to the parafermionic partition
function can be written as;
\begin{eqnarray}
\label{toto26} \frac{
\Delta_s^{+}(\beta)}{Z_{s}(\beta)}&=&\frac{\prod_{k=1}^{\infty}(1-
x^{k} + x^{2k}- \cdots - x^{(s-1)k})}{\prod_{k=1}^{\infty}(1+
x^{k} + x^{2k}+ \cdots +
x^{(s-1)k})}\nonumber\\
&=&\prod_{k=1}^{\infty}\frac{(1-x^k)}{({1+x^k})},\;\mbox\;s\;\mbox{even}
\end{eqnarray}

\begin{eqnarray}
\label{toto27}
 \frac{
\Delta_s^{-}(\beta)}{Z_{s}(\beta)}&=&\frac{\prod_{k=1}^{\infty}(1-
x^{k} + x^{2k}- \cdots + x^{(s-1)k})}{\prod_{k=1}^{\infty}(1+
x^{k} + x^{2k}+ \cdots +
x^{(s-1)k})}\nonumber\\
&=&\prod_{k=1}^{\infty}\frac{(1-x^k)}{({1+x^k})}\frac{(1+x^{sk})}{(1-x^{sk})},\;\mbox\;s\;\mbox{odd}.
\end{eqnarray}
Here, we would like to make some remarks  about the above
identities and their implications. The first identity shows that
for all $s$ even, the ratio of the graded parafermionic partition
function to the parafermionic partition function is always given
by $\prod_{k=1}^{\infty}\frac{(1-x^k)}{({1+x^k})}$. This ratio is
known to be equal to $\theta_4(0,x)=\sum_{n=-\infty}^\infty (-1)^n
x^{n^{2}}$ via Gauss's identity. The second identity, in the case
of $s$ odd, does not give  $\theta_4(0,x)$ but rather the ratio
$\theta_4(0,x)/\theta_4(0,x^s)$ so this extra factor can be
thought of as correction factor that one has to insert in $\frac{
\Delta_s^{+}(\beta)}{Z_{s}(\beta)}$ to obtain the odd case. Now by
combining the above identities we obtain the following identity,

\begin{eqnarray}
\label{toto28} \frac{\Delta_s^{-}(\beta)}{Z_{s}(\beta)}
&=&\frac{\Delta_{2}^{+}(\beta)}{Z_{2}(\beta)}
\prod_{k=1}^{\infty}\frac{(1+x^{sk})}{(1-x^{sk})}\nonumber\\
&=&\frac{\theta_4(0,x)}{\theta_4(0,x^s)},\;\mbox\;s\;\mbox{odd}.
\end{eqnarray}
Now, let us consider the graded parafermion of order three. Its
partition function is $\Delta_3^{-}(\beta)=\prod_{k=1}^{\infty}(1-
x^{k} + x^{2k})=\prod_{k=1}^{\infty}\frac{(1+x^{3k})}{({1+x^k})}$,
and as for the parafermion of order three the right hand side of
this identity can be simplified to give
\begin{eqnarray}
\label{toto29}
\Delta_{3}^{-}(\beta)&=&\prod_{k=1}^{\infty}(1-
x^{k}+
x^{2k})\nonumber\\
&=&\prod_{k=1}^{\infty}\frac{(1+x^{3k})}{({1+x^k})}\nonumber\\
&=&\prod_{k=1}^{\infty}\frac{1}{(1+x^{3k-2})}\frac{1}{(1+x^{3k-1})}.
\end{eqnarray}
The inverse of this partition function is nothing but the
generating function of the number of partitions of $k$ into
distinct parts which are prime to 3. We will see below that the
generating function of Euler on partitions without restrictions is
obtained by inverting the graded fermionic partition function of
order two. In 1926, I.Schur \cite{schur} proved the following
theorem
\begin{theorem}
{ the number of partitions of $k$ into distinct parts which are
prime to 3 is identical with the number of partitions of a $k$
into parts congruent to 1 or 5 modulo 6, in addition  both of
these number of partitions are equal to the number of partitions
of $k$ of the form $ b_1+b_2+\cdots +b_l$ such that
$b_i-b_{i+1}\geq3$ with strict inequality if $b_i$ is a multiple
of 3}.
\end{theorem}
Therefore as a result we obtain the following identity,
\begin{equation}
\label{toto30} \frac{1}{\Delta_{3}^{-}(\beta)}=
\prod_{k=1}^{\infty}\frac{1}{(1-x^{6k-5})}\frac{1}{(1-x^{6k-1})},
\end{equation}
so we learn that the inverted graded parafermionic partition
function of order 3 coincides with the generating function for the
numbers of partitions in the Schur's theorem. In the following we
would like to make a connection between the graded parafermion of
even order, say $s=2l$, $l$ any positive integer and the graded
parafermion of order $l$. Since the term $(1+(-1)^{2l-1}x^{2lk})$
in the expression of the graded parafermionic partition function
factorizes and hence the expression relating the inverted
parafermionic partition functions of order $2l$ to that of order
$l$ is,
\begin{eqnarray}
\label{toto31} \frac{1}{\Delta_{2l}^{+}(\beta)}&=&
\prod_{k=1}^{\infty}\frac{(1+x^{k})}{(1+(-1)^{2l-1}x^{2lk})}\nonumber\\
&=&\prod_{k=1}^{\infty}\frac{(1+x^{k})}{(1+(-1)^{l-1}x^{lk})}\frac{1}{(1+(-1)^{l}x^{lk})}.
\end{eqnarray}
The parafermionic partition function of order one is the Witten
index
\begin{eqnarray}
\Delta_1^{-}(\beta)={\rm
Tr}\left[(-)^F\exp{-\beta(H_B+H_F)}\right]=1,\nonumber
\end{eqnarray}
so both the the graded and the inverted parafermionic partition
functions are equal to 1. Therefore the inverted graded
parafermionic partition function of order two is,
\begin{eqnarray}
\label{toto32}
 \frac{1}{\Delta_{2}^{+}(\beta)}&=&
\prod_{k=1}^{\infty}\frac{(1+x^{k})}{(1-x^{2k})}\nonumber\\
&=&1\prod_{k=1}^{\infty}\frac{1}{(1-x^{k})},
\end{eqnarray}
which is exactly the Euler generating function for the
unrestricted partitions $p(k)$ of a number $k$ that we mentioned
above. The order 4 inverted parafermionic partition function
should be related to the Euler generating function,
\begin{eqnarray}
\label{toto33}
 \frac{1}{\Delta_{4}^{+}(\beta)}&=&
\prod_{k=1}^{\infty}\frac{(1+x^{k})}{(1-x^{4k})}\nonumber\\
&=&\prod_{k=1}^{\infty}\frac{1}{(1-x^{k})}\frac{1}{(1+x^{2k})} \nonumber\\
&=&\prod_{k=1}^{\infty}\frac{(1+x^{2k-1})}{(1-x^{2k})},
\end{eqnarray}
where the right hand side of this equation gives the number of
partitions of $k$ in which each even part occurs with even
multiplicity and there is no restriction on the odd distinct
parts. the next generating function we look for is the one
connected with the Schur theorem  and it is the inverted sixth
order parafermionic partition function,
\begin{eqnarray}
\label{toto34}
 \frac{1}{\Delta_{6}^{+}(\beta)}&=&
\prod_{k=1}^{\infty}\frac{(1+x^{k})}{(1-x^{6k})}\nonumber\\
&=&\prod_{k=1}^{\infty}\frac{(1+x^{3k-1})(1+x^{3k-2})}{(1-x^{3k})}.
\end{eqnarray}
This gives the number of partitions of $k$ in which the distinct
parts are prime to 3 and the unrestricted parts contain 3 and its
multiples. Using Maple one could generate sequences of partitions
from our formulae depending on $s$. When $s=3$ we have,
\begin{center}$1+x+x^{2}+\cdots+3x^{7}+3x^{8}+3x^{9}+4x^{10}+\cdots $\end{center}
so for example the number of distinct partitions of the number 10
in which the parts are prime to 3 is 4 (which we denote by
$a(10)=4$) because $10=8+2=7+2+1=5+4+1$. The corresponding
sequence of partitions is,\begin{center} $
1,1,1,1,1,2,2,3,3,3,4,5,6,7,8,9,10,12,\cdots $,\end{center} this
corresponds to the sequence $A003105$ on the on-Line Encyclopedia
of integer sequences \cite{sloane}. The next sequence of
partitions of the number $k$ would be that which corresponds to
$s=4$, which is,\begin{center} $1,1,1,2,3,4,5,7,10,13,16,21,\cdots
$\end{center} this is also in \cite{sloane} and corresponds to the
sequence $A006950$. It was pointed out by Sellers that the number
of partitions  for this sequence that we mentioned above for $s=4$
it is also the number of partitions of $k$ into parts not
congruent to 2 mod 4. so for example number of partition of 7
equals 7, $a(7)=7$
because\begin{center}$7=5+1+1=4+3=4+1+1+1=3+3+1=3+1+1+1+1=1+1+1+1+1+1+1$.\end{center}
The fifth and the sixth orders from our formulae for the inverted
parafermionic partition function give the following sequences of
partition of the number $k$,
\begin{center}
$1,1,1,2,2,2,3,4,4,6,7,8,10,16.19,22,26,30,35,41,\cdots$\end{center}
for the fifth order and
\begin{center}$1,1,1,2,2,3,5,6,7,10,12,15,25,30,39,46,\cdots$\end{center}
for the sixth order respectively. These two sequences  are new
sequences and now they appear on the on-Line Encyclopedia of
integer sequences \cite{sloane} with references $A096938$ and
$A096981$. In the fifth order inverted parafermionic partition
function,the number 5 and its multiples are not present in the
number of partitions of $k$ into distinct parts, for example
$8=7+1=6+2=4+3+1$ so $a(8)=4$ also $9=8+1=7+2=4+3+2$ then
$a(9)=4$. For the sixth order sequence the number of partitions of
the number 11 for example, is 15, $a(11)=15$.Because
\begin{center}$
11=10+1=8+2+1=7+4=5+4+2=9+2=8+3=7+3+1=6+5=6+4+1=6+3+2=5+3+2+1=4+3+3+1=3+3+3+2.$\end{center}
Therefore, we see that our generating function gives known
sequences for partitions of numbers and as well as new ones. We
would like to make a comment about our generating function, for
example in the third order see the left hand side of equation
(\ref{toto30}) this was given in the sequence A00305. At the
fourth order, however, a formula like ours was not given in the
sequence A006950 see \cite{sloane}.

Finally we mention another new sequence that we have found in this
work and which follows from  the identity in equation
(\ref{toto28}). Setting $s=3$ then we have
\begin{equation}
\frac{Z_{3}(\beta)}{\Delta_3^{-}(\beta)}=
\frac{\theta_4(0,x^3)}{\theta_4(0,x)},\nonumber
\end{equation}
and using Maple the sequence
is,\begin{center}$1,2,4,6,10,16,24,36,52,74,104,144,198,268,360,\cdots,48672,59122,\cdots$.\end{center}
This corresponds to the number of partitions of $2k$ prime to 3
with all odd parts occurring with even multiplicities. There is no
restriction on the even parts e.g $a(8)=10$ because
\begin{center}
$8=4+4=4+2+2=2+2+2+2=2+2+2+1+1=2+2+1+1+1+1=2+1+1+1+1+1+1=4+1+1+1+1=1+1+1+1+1+1+1+1.$\end{center}
This new sequence has reference number A098151 in the on-Line
Encyclopedia of integer sequences \cite{sloane}. From the identity
given by equation (\ref{toto27}) for $s$ even one has,
\begin{equation}
\frac{Z_{s}(\beta)}{\Delta_s^{+}(\beta)}=
\frac{1}{\theta_4(0,x)},\nonumber
\end{equation}
this generates a sequence which coincides with the sequence
A0151128 in \cite{sloane}, using Maple we obtain
\begin{center}$1,2,8,14,24,40,64,100,154,\cdots,9904,13288,\cdots,$\end{center}
which is the number of partitions of $2k$ with all odd parts
occurring with even multiplicities. There is no restriction on the
even parts.

\section{CONCLUSIONS}
In this paper partial supersymmetry was used to derive Euler
theorem from a fermionic partition function and from the
parafermionic partition function a theorem in the theory of
partitions, which says that the number of partitions of $k$ in
which no parts appear more than $s-1$ times equals the number of
partitions of $k$ such that no parts is divisible by $s$. In the
last section we obtained the expression for the graded
parafermionic partition function and we saw that the inverted
parafermionic partition function generates partition of numbers
with  given restrictions. The generating function we have obtained
is general in the sense that when the order of the graded
parafermion is two our generating function coincides with that of
the Euler generating function of partitions without restrictions.
We have  also shown that these generating functions are related to
each other, the order $2l$ is expressed in terms of order $l$
where $l$ is any  positive integer. Also in the last section we
have obtained new sequences of partitions \cite{sloane}. The
theory used here may be called additive quantum theory as the
bosonic partition function is the Euler generating function for
the unrestricted partitions. The Euler theorem  equates the number
of distinct partitions with the number of unrestricted odd
partitions.The generating function proof of this theorem is simple
in number theory and this is also the case using partial
supersymmetry. One interpretation we give is that the fermionic
partition function is the odd part of the bosonic partition
function. One may also look at it differently and write
$Z_f(\beta)=\Delta_{F}(2\beta)/\Delta_{F}(\beta)$ from which we
learn that $\Delta_{F}(2\beta)=Z_f(\beta)\Delta_{F}(\beta)$ so
mixing the fermionic system with the graded fermionic system at
thermal equilibrum at one temperature is the same as a graded
parafermionic system with different temperature.This also happens
in the case of quantum field theory with a logarithmic spectrum
\cite{spector} in which the term duality was used to characterize
the identities among arithmetic quantum theories. In our case the
latter formula generalizes naturally by simply using equation
(\ref{toto16}) or equations (\ref{toto26}) and (\ref{toto27})
which read, $\Delta_F(s\beta)=Z_s(\beta)\Delta_{F}(\beta)$ for
both $s$ even and odd, this identity will be a trivial identity
when s goes to infinity as both bosonic and graded fermionic
partition functions cancel each other and we recover the Witten
index which is a complete cancellation between boson and fermions.
An other way to look at the equations (\ref{toto26}) and
(\ref{toto27}) is that in the former equation the ratio of the
graded parafermionic partition function to the parafermionic
partition function when $s$ is even is always given by
$\theta_4(0,x)$. When $s$ is odd, however, it is given by the
ratio $\theta_4(0,x)/\theta_4(0,x^s)$.Finally, the new sequence
A096981 for the sixth order that I have obtained is also
equivalent to the number of partitions of $k$ into parts congruent
to {0,1,3,5} mod 6 see sequence A096981 in \cite{sloane} for
details on the sequence and its connection with other sequences.\\

\vspace{7mm} {\bf Akcknowledgment:}

I would like to thank M.O'Loughlin, G.Thompson, A.S.Verjovsky for
critical reading of the manuscript and discussions, D.Spector for
correspondence and the Abdus Salam  ICTP, and  SISSA for support
and hospitality throughout these years.


\newpage

\bibliographystyle{phaip}

\begin{thebibliography}{1}

\bibitem{donald}
B. Julia, J. Phys. (France) {\bf 50},1371 (1989); \\
D .Spector, Phys. Lett. A {\bf 140},311 (1989);\\
D. Spector,comm. Math. Phys. \textbf{127},239 (1990).


\bibitem{i.bakas}
I. Bakas and M. Bowiwick,J. Math. Phys. \textbf{32},1881 (1991).

\bibitem{spector}
D. Spector,J. Math. Phys. \textbf{32},1919 (1998).

\bibitem{apostol}
T. M. Apostol,An introduction to Analytic Number
Theory(Springer-Verlag,New-York,1976).

\bibitem{andrews} G. E . Andrews,The Theory Of Partitions,(Addison-Wesley,1976).

\bibitem{sellers}
J. A. Sellers Journal of Integer Sequences. \textbf{7} (2004)
Article 04.2.4

\bibitem{sloane}
N. J. A.Sloane,The On-Line Encyclopedia of Integer
Sequences.Published electronically at
http://www.research.att.com/$\sim$njas/sequences/.

\bibitem{witten}
E. Witten,Nucl. Phys. B{\textbf 202}253 (1982).

\bibitem{schur}
I. Schur,Gesmmelte Abhandlungen. (Springer-Verlag,Berlin,{\bf
12}43 1973).


\end{thebibliography}

\end{document}